\documentclass[usenatbib,useAMS,a4paper]{mn2e}

\usepackage{txfonts}
\usepackage{epsfig}
\usepackage{lscape}

\def\msun{{\rm M_{\odot}}}

\title [Multiple ionisation sources in H~{\sc ii} regions]{Multiple ionisation 
sources in H~{\sc ii} regions and their effect on derived nebular abundances.}
\author[Ercolano, Wesson \& Bastian]{B. Ercolano$^{1,2}$, R. Wesson$^{2}$ and N. Bastian$^{1}$\\
$^1$Institute of Astronomy, Madingley Rd, Cambridge, CB3 0HA, UK \\
$^2$Department of Physics and Astronomy, University College London, WC1E 6BT, UK}

\date{Submitted:}

\pagerange{\pageref{firstpage}--\pageref{lastpage}} \pubyear{}

\begin{document}
\def\lta{\mathrel{\spose{\lower 3pt\hbox{$\mathchar"218$}}
     \raise 2.0pt\hbox{$\mathchar"13C$}}}
\def\gta{\mathrel{\spose{\lower 3pt\hbox{$\mathchar"218$}}
     \raise 2.0pt\hbox{$\mathchar"13E$}}}
\def\Msun{{\rm M}_\odot}
\def\msun{{\rm M}_\odot}
\def\Rsun{{\rm R}_\odot}
\def\Lsun{{\rm L}_\odot}
\def\19{GRS~1915+105}
\label{firstpage}
\maketitle

\begin{abstract}

We present a theoretical investigation of the effect of multiple
ionisation sources in H~{\sc ii} regions on the
total elemental abundances derived from the analysis of
collisionally excited emission lines. We focus on empirical methods
based on direct temperature measurements that are commonly employed in
cases when the temperature of the nebular gas can be determined from
the ratio of nebular to auroral lines of (e.g.) doubly ionised
oxygen. We find that direct temperature methods that employ a
two-temperature zone approach (DT2T methods) are very robust against
the spatial distribution of 
sources. Errors smaller than 0.15 dex are estimated for regions where
the metallicity is twice solar and errors below 0.05 dex for solar
metallicities and below. The biases introduced by the spatial
distribution of the ionisation sources are thus much smaller for DT2T
methods than for strong line methods, previously investigated by
Ercolano, Bastian \& Stasi\'nska. Our findings are in agreement with
the recent study of H~{\sc ii} regions in NGC~300 by Bresolin et al. 

\end{abstract}


\section{Introduction}

The analysis of emission lines from H~{\sc ii} regions powered by OB stars is
often the only means available for the determination of gas abundances
both in our Galaxy and others. Accurate abundance determinations are
crucial to derive metallicity gradients, which provide a key
observational constraint to chemical evolution models of galaxies.  
The spectra of H~{\sc ii} regions in the optical and infrared are dominated
by collisionally excited lines (CELs) of singly and doubly ionised
ions of some of the more abundant heavy elements (e.g. oxygen, carbon etc.). While
the metallicity diagnostic power of CELs from H~{\sc ii} regions is widely
recognised and used for studying the chemical composition of galaxies,
the uncertainties inherent to the empirical methods employed are often
overlooked, potentially leading to somewhat over-optimistic error
estimates.  

Ionic abundances can be obtained from emission lines via the solution
of the statistical equilibrium equation under the assumption of a gas
temperature and density. Both gas temperature and density can be
empirically derived from the observations using the ratio of
diagnostic emission lines (e.g. Osterbrock and Ferland, 2006). A widely used
temperature diagnostic relies on the ratio of nebular to auroral
[OIII] lines, in particular
([OIII]$\lambda$5007+[OIII]$\lambda$4959)/[OIII]$\lambda$4363. This
method is often referred to as the 'direct temperature' method. For
faint (distant) or metal-rich regions, however, the
[OIII]$\lambda$4363 line is seldom detected, rendering impossible the
direct determination of temperature. In these cases one is often
forced to use 'strong-line methods', which rely on ratios of some of
the strongest CELs calibrated with one-dimensional photoionisation
models.  In a previous paper (Ercolano, Bastian \& Stasi\'nska, 2007,
EBS07) we showed that commonly employed strong-line methods, may
introduce a {\it systematic} bias of typically 0.1-0.3~dex, but up to
1~dex on the logarithmic oxygen abundance for regions where the gas
and the ionising stars are intermixed.  

Direct-temperature methods should be more reliable than strong line
methods; however even these are not entirely immune by errors
introduced by calibrations from spherically symmetric models. While
the error on the determination of a single ionic abundance from
emission lines via the solution of the statistical equilibrium matrix is
simply determined by the accuracy of the gas temperature and (to a
lesser extent) density estimates, not all the ionic stages of a given
element produce emission lines in the observed wavelength range. For
this reason the determination of total elemental abundances from ionic
abundances relies on a correction for the unseen stages of ionisation,
known as the ionisation correction factor (ICF). A potential bias is
therefore introduced by the ICF schemes that are themselves calibrated
via one-dimensional photoionisation modeling (e.g. Kingsburgh \&
Barlow 1994, Peimbert et al., 1992), with the implied assumption of a
geometry comprising of a spherical region with a single central
location for all ionisation sources.   

The aim of this paper is to estimate whether the use of ICF schemes to
derive total elemental abundances of H~{\sc ii} regions from direct-temperature
methods introduces a bias in the case of regions ionised by
multiple stars intermixed with the nebular gas. Via the
analysis of theoretical H~{\sc ii} region spectra obtained with the {\sc mocassin} code (Ercolano et al. 2003, 2005, 2008) in the set-up used by EBS07,
we show that, unlike strong line methods, direct temperature methods
using a two-temperature approach are very reliable also in cases when
the stars are fully distributed within the nebular gas. These results
are in line with the recent findings of Bresolin et al (2009, B09) who
found good agreement between CEL abundances obtained by direct
temperature methods with the abundances derived from absorption line
analysis of stellar photospheres in a sample of H~{\sc ii} regions in the
spiral galaxy NGC~300.  

The paper is organised as follows. Section 2 summarises the model
setup and input parameters. Section 3 contains a brief description of
the methods employed. Our results are given in Section 4, while
Section 5 is dedicated to a discussion and conclusions.  

\section{Theoretical Emission Line Spectra: Model Setup and Input Parameters}
\label{s:m}
We have used the theoretical nebular spectra of EBS07, which were
obtained using the three-dimensional photoionisation code {\sc mocassin}
(Ercolano et al. 2003, 2005, 2008). This code uses a Monte Carlo
approach to the transfer of radiation and can easily deal with
multiple ionisation sources arbitrarily distributed within the
simulation region. The atomic database included opacity data from
Verner et al. (1993) and Verner \& Yakovlev (1995), 
energy levels, collision strengths and transition probabilities from
Version 5.2 of the CHIANTI database (Landi et al. 2006, and references
therein) and the improved hydrogen and helium free-bound continuous
emission data of Ercolano \& Storey (2006).  The model setup and input
parameters are described 
in EBS07, here we summarise briefly the main points, but refer the
interested reader to EBS07 for further details.  

The regions are assumed to be spherical and consisting of homogeneous
gas with number density N$_H$ = 100~cm$^{-3}$. The total number of
ionising photons is constant for all models and is Q$_{\rm H^{\rm 0}}
= 2.8 \times 10^{50}$~s$^{-1}$. We consider models of five different
metallicities ( Z/Z$_{\odot}$= 0.05, 0.2, 0.4, 1.0 and 2.0). The solar
abundance model assumes the values of Grevesse and Sauval (1998) with
the exception of C, N and O abundances which are taken from Allende
Prieto et al. (2002), Holweger (2001) and Allende Prieto, Lambert and
Asplund (2001), respectively. The higher and lower metallicity cases
were obtained from the solar abundances by scaling using the empirical
abundance trends observed in H~{\sc ii} regions by Izotov et al (2006). The
gas is ionised by 240 sources belonging to two populations, a
hot (M$_*$ = 56 M$_{\odot}$) and cool (M$_*$ = 37 M$_{\odot}$)
population, each population, as a whole, emits equal quantities
of H-ionising photons.  
The stellar spectra were computed with the {\sc starburst99} spectral
synthesis code (Leitherer et al. 1999) with the up-to-date non-LTE
stellar atmospheres implemented by Smith, Norris and Crowther (2002)
using single isochrones for the appropriate stellar masses. The stars
were distrubuted as follows: (i) centrally concentrated at the centre
of the spherical region - C-models; (ii) distributed in the
half-volume of the spherical region - H-models; (iii) distributed in
the full spherical volume - F-models. In the C-models all stars
  share the same location at the origin of the Cartesian axes. In the
  H- and F-models the stars are distributed stochastically such as to
  obtain a statistically homogeneous 3D distribution of sources in the
  half or full spherical region, respectively. The Stroemgren sphere of stars
in the F-models seldom overlap, while those of the C-model completely
overlap, with the H-model representing the intermediate case. 

In Table~1 we list the subset of the emission lines from our
theoretical spectra that were used for the analysis described in the
following section.

\section{The direct temperature method}

The aim of this study is to test whether the abundances determined
from the emission  
line spectrum of an H~{\sc ii} region ionised by multiple stars intermixed
with the gas are different from the abundances determined by a region
with exactly the same physical characteristics but with all ionisation
sources concentrated in the centre. EBS07 indeed demonstrated that
large biases are introduced by this effect when strong-line methods
are used. The main reason for such differences was due to a decrease
of the 'effective ionisation parameter' of the gas when the stars were
fully distributed within the medium compared to when the same stars
were all concentrated at the centre of the nebula. This affected the
temperature structure of the nebula significantly enough to produce
large errors in the derived abundances.  

Ionic abundances from direct-temperature methods should be immune from
this error as long as the temperature gradients within a given ionic
phase are not too large (see e.g. Stasi\'nska, 1980 and Kingdon \&
Ferland, 1995). However total elemental abundances can only be
obtained by applying an ICF scheme to correct for the unseen ionisation
stages. The question therefore remains as to what is the effect
of the geometrical distribution of the stars on the ICFs, which rely
on theoretical calibrations via one-dimensional photoionisation models
(e.g. Kingsburgh \& Barlow, 1994, hereafter KB94; Peimbert et al.,
1992, hereafter PG92).  

To answer this question we took a subset of lines typically
observed in extragalactic H~{\sc ii} regions, including the important
temperature diagnostic lines of [O~{\sc iii}] at 4363{\AA} and [N~{\sc
    ii}] at 5754{\AA}, from the line spectra produced by the 
photoionisation models described in Section 2. We then used these
model spectra to derive chemical abundances via the direct temperature
method.  The
lines used for the analysis are listed in Table~1. 

We considered the nebula as being composed of two separate zones of low and high ionisation: first, we assumed a temperature of
10\,000\,K to obtain electron densities from the [O~{\sc ii}]
$\lambda$3726/$\lambda$3729 and [S~{\sc ii}]
$\lambda$6717/$\lambda$6731 line ratios.  Then, the average of these
two electron densities was used to derive a temperature from the
[N~{\sc ii}] ($\lambda$6548+6584)/$\lambda$5754 line ratio.  This
temperature was then used to recalculate the densities, and the
temperature recalculated once more using the resulting density.  The
abundances of singly-ionised species were derived using this
temperature and density. 
Then, we used the same iterative approach, but using the [Cl~{\sc
    iii}] 5517/5537 and [Ar~{\sc iv}] 4711/4740 line ratios as density
diagnostics, and the [O~{\sc iii}] 4959+5007/4363 ratio as a
temperature diagnostic.  Abundances of doubly and more highly ionised
species were derived using this temperature and density. 

Total abundances relative to hydrogen were calculated for He, C, N, O,
Ne, S and Ar, using the two commonly used ICF schemes of KB94 and PTPR92.
For several atoms these schemes use the same correction; they differ
for helium (KB94 does not correct for neutral helium while PTPR92
does), argon and sulfur. 

\begin{landscape}
\begin{table}

\caption{Model emission line fluxes, relative to H$\beta$=100.} \label{t:t1}

\resizebox{23cm}{!}{

\begin{tabular}{lllllllllllllllll}
\hline
$\lambda$ (\AA) & Ion & \multicolumn{15}{c}{Model}\\
 & & 0.05c & 0.05f & 0.05h & 0.20c & 0.20f & 0.20h & 0.40c & 0.40f & 0.40h & 1.00c & 1.00f & 1.00h & 2.00c & 2.00f & 2.00h \\ 
\hline
3726.03 & [O~{\sc ii}]  &12.38 &27.70 &14.07 &40.05 &87.69 &35.55 &35.20 &95.33 &32.05 &18.55 &48.28 &17.57 &63.89 &62.60 &55.92\\
3728.82 & [O~{\sc ii}]  &16.51 &37.10 &18.78 &53.18 &116.7 &47.19 &46.62 &126.5 &42.47 &24.34 &63.43 &23.04 &82.17 &80.68 &71.92\\
3868.75 & [Ne~{\sc ii}] &22.93 &11.58 &21.12 &43.07 &26.43 &44.01 &31.03 &21.10 &31.42 &2.682 &2.433 &2.410 &0.034 &0.149 &0.040\\
3967.46 & [Ne~{\sc ii}] &6.910 &3.491 &6.363 &12.97 &7.965 &13.25 &9.349 &6.359 &9.468 &0.808 &0.733 &0.726 &0.010 &0.045 &0.012\\
4068.60 & [S~{\sc ii}]  &0.471 &1.828 &0.634 &1.426 &5.011 &1.341 &1.590 &6.435 &1.547 &1.020 &4.000 &0.950 &0.351 &1.072 &0.331\\
4076.35 & [S~{\sc ii}]  &0.163 &0.632 &0.219 &0.493 &1.734 &0.464 &0.550 &2.227 &0.535 &0.354 &1.390 &0.329 &0.122 &0.374 &0.115\\
4363.21 & [O~{\sc iii}] &9.693 &3.995 &8.741 &8.649 &4.593 &9.138 &2.993 &1.747 &3.037 &0.070 &0.045 &0.053 &0.002 &0.005 &0.002\\
4471.50 & He~{\sc i}    &4.150 &4.186 &4.143 &4.328 &4.241 &4.305 &4.469 &4.295 &4.460 &4.723 &4.476 &4.723 &4.293 &4.356 &4.345\\
4685.68 & He~{\sc ii}   &0.008 &0.011 &0.010 &0.013 &0.013 &0.013 &0.007 &0.010 &0.009 &0.006 &0.009 &0.006 &0.003 &0.001 &0.002\\
4711.37 & [Ar~{\sc iv}] &1.385 &0.491 &1.235 &2.319 &0.898 &2.451 &1.765 &0.710 &1.890 &0.081 &0.047 &0.082 &0.000 &0.002 &0.001\\
4740.17 & [Ar~{\sc iv}] &1.069 &0.377 &0.952 &1.737 &0.672 &1.838 &1.295 &0.520 &1.386 &0.059 &0.034 &0.060 &0.000 &0.001 &0.000\\
4861.33 & H~{\sc i}     &100.0 &100.0 &100.0 &100.0 &100.0 &100.0 &100.0 &100.0 &100.0 &100.0 &100.0 &100.0 &100.0 &100.0 &100.0\\
4958.91 & [O~{\sc iii}] &110.1 &52.62 &101.5 &200.4 &116.4 &207.7 &161.3 &102.3 &165.2 &19.47 &16.55 &17.97 &1.854 &4.190 &1.844\\
5006.84 & [O~{\sc iii}] &328.6 &157.0 &303.0 &598.0 &347.4 &619.8 &481.5 &305.2 &493.0 &58.11 &49.40 &53.64 &5.531 &12.50 &5.504\\
5754.60 & [N~{\sc ii}]  &0.026 &0.059 &0.030 &0.076 &0.165 &0.068 &0.097 &0.258 &0.089 &0.075 &0.193 &0.071 &0.237 &0.266 &0.215\\
5875.66 & He~{\sc i}    &10.87 &11.16 &10.87 &11.57 &11.42 &11.50 &12.27 &11.80 &12.23 &14.32 &13.22 &14.26 &14.12 &13.42 &14.15\\
6312.10 & [S~{\sc iii}] &0.923 &1.045 &0.982 &2.261 &2.867 &2.142 &1.868 &2.573 &1.686 &0.458 &0.743 &0.410 &0.368 &0.309 &0.320\\
6548.10 & [N~{\sc ii}]  &0.321 &0.943 &0.396 &1.114 &2.919 &1.017 &1.895 &5.979 &1.784 &2.831 &9.155 &2.741 &16.14 &21.44 &15.15\\
6562.77 & H~{\sc i}     &276.8 &280.8 &277.3 &281.2 &283.0 &281.1 &286.8 &287.1 &286.7 &306.5 &301.3 &305.7 &314.4 &306.6 &313.7\\
6583.50 & [N~{\sc ii}]  &0.981 &2.880 &1.209 &3.402 &8.916 &3.107 &5.787 &18.26 &5.449 &8.647 &27.96 &8.371 &49.31 &65.50 &46.28\\
6678.16 & He~{\sc i}    &2.983 &3.115 &2.993 &3.238 &3.214 &3.218 &3.486 &3.354 &3.476 &4.080 &3.784 &4.069 &3.864 &3.792 &3.905\\
6716.44 & [S~{\sc ii}]  &4.194 &17.80 &5.839 &12.84 &50.82 &12.43 &16.03 &70.59 &15.77 &13.80 &61.42 &13.08 &6.061 &21.11 &5.783\\
6730.82 & [S~{\sc ii}]  &3.100 &13.11 &4.314 &9.524 &37.39 &9.199 &11.83 &51.96 &11.64 &10.15 &45.06 &9.619 &4.447 &15.50 &4.242\\
7136.80 & [Ar~{\sc iii}] &0.808 &1.193 &0.875 &2.969 &4.414 &2.833 &3.662 &5.865 &3.412 &2.342 &4.033 &2.205 &3.014 &4.011 &2.592\\
7319 & [O~{\sc ii}]  &0.544 &1.025 &0.597 &1.479 &2.856 &1.290 &1.015 &2.393 &0.902 &0.311 &0.676 &0.291 &0.644 &0.568 &0.551\\
7330 & [O~{\sc ii}]  &0.443 &0.836 &0.487 &1.205 &2.329 &1.052 &0.828 &1.953 &0.736 &0.254 &0.552 &0.238 &0.526 &0.464 &0.450\\
7751.43 & [Ar~{\sc iii}] &0.194 &0.286 &0.210 &0.711 &1.058 &0.679 &0.877 &1.405 &0.818 &0.561 &0.966 &0.528 &0.722 &0.961 &0.621\\
\hline
\end{tabular} 
}
\end{table} 
\end{landscape}

We also considered the case where a low-ionisation temperature
diagnostic is not available.  In this situation, one can either use
the [O~{\sc iii}] temperature for all ions, or estimate the low
ionisation temperature using relations such as that found by Pilyugin
et al. (2006).  We confirm previous findings that applying a single
temperature diagnostics introduces considerable errors into the
abundance determinations. This case will not be further discussed.

\begin{figure}
\begin{center}
\includegraphics[width=9cm]{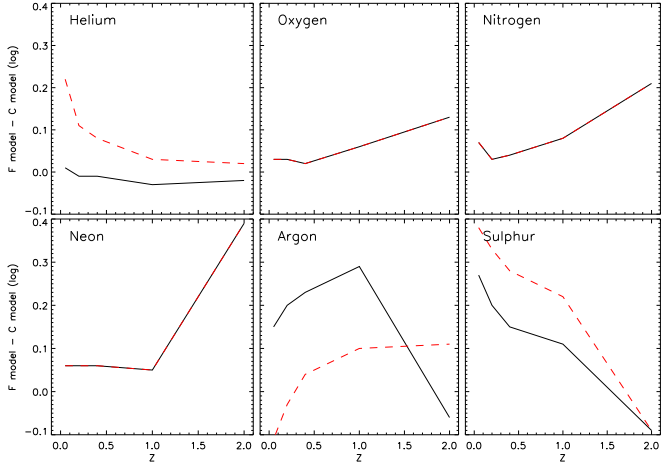}
\caption[]{Error on the logarithmic abundances calculated via the DT2T
  method caused by the spatial distribution of ionisation sources. The
  black solid line shows results obtained via the KB94 ICFs, while the
  red dashed line shows those obtained via the PTR92 ICFs.} 
\label{f:f1}
\end{center}
\end{figure}

\begin{table*}
\caption[]{Logarithmic errors and ICF corrections}
\label{t:t1}
\begin{center}
\begin{tabular}{lcccc|lcccc|lcccc}
\hline
\multicolumn{5}{|c|}{Helium} &\multicolumn{5}{|c|}{Oxygen} &\multicolumn{5}{|c|}{Nitrogen} \\ 
 Z/Z$_{\odot}$  & \multicolumn{2}{|c|}{$E_{F-C}$} & \multicolumn{2}{|c|}{$\Delta(E_{F-C})$} &  Z/Z$_{\odot}$  & \multicolumn{2}{|c|}{$E_{F-C}$} & \multicolumn{2}{|c|}{$\Delta(E_{F-C})$}   & Z/Z$_{\odot}$  & \multicolumn{2}{|c|}{$E_{F-C}$} & \multicolumn{2}{|c|}{$\Delta(E_{F-C})$} \\
               &    KB94    &  PTR92         &    KB94    &  PTR92                          &                 &    KB94    &  PTR92         &    KB94    &  PTR92                          &                  &    KB94    &  PTR92         &    KB94    &  PTR92                      \\   
\hline                                                                                         \hline                                                                                          \hline                                                                                       
0.05           &    0.01    &   0.22         &   1.0      &  -0.14                          &  0.05           & 0.03       &  0.03          &    1.0     &   1.0                           &   0.05           &      0.07  &  0.07          &    -0.03   &    -0.03\\   
0.2            &   -0.01    &   0.11         &   1.0      &  -0.06                          &  0.2            & 0.03       &  0.03          &    1.0     &   1.0                           &   0.2            &      0.03  &  0.03          &    +0.08   &    +0.08\\   
0.4            &   -0.01    &   0.08         &   1.0      &  -0.03                          &  0.4            & 0.02       &  0.02          &    1.0     &   1.0                           &   0.4            &      0.04  &  0.04          &    +0.03   &    +0.03\\   
1.0            &   -0.03    &   0.03         &   1.0      &  -0.007                         &  1.0            & 0.06       &  0.06          &    1.0     &   1.0                           &   1.0            &      0.08  &  0.08          &    -0.03   &    -0.03\\   
2.0            &   -0.02    &   0.02         &   1.0      &  -0.047                         &  2.0            & 0.13       &  0.13          &    1.0     &   1.0                           &   2.0            &      0.21  &  0.21          &    -0.14   &    -0.14\\   
\hline
\multicolumn{5}{|c|}{Neon} &\multicolumn{5}{|c|}{Argon} &\multicolumn{5}{|c|}{Sulfur} \\ 
 Z/Z$_{\odot}$  & \multicolumn{2}{|c|}{$E_{F-C}$} & \multicolumn{2}{|c|}{$\Delta(E_{F-C})$} &  Z/Z$_{\odot}$  & \multicolumn{2}{|c|}{$E_{F-C}$} & \multicolumn{2}{|c|}{$\Delta(E_{F-C})$}   & Z/Z$_{\odot}$  & \multicolumn{2}{|c|}{$E_{F-C}$} & \multicolumn{2}{|c|}{$\Delta(E_{F-C})$} \\
               &    KB94    &  PTR92         &    KB94    &  PTR92                          &                 &    KB94    &  PTR92         &    KB94    &  PTR92                          &                  &    KB94    &  PTR92         &    KB94    &  PTR92                      \\   
\hline                                                                                         \hline                                                                                          \hline                                                                                       
0.05           &     0.06   &       0.06     &  -0.07     &     -0.07                       &  0.05           &    0.15    &      -0.11     &  -0.07     &        1.0                      &   0.05           &     0.27   &     0.38      &     -0.21   &  -0.33                  \\   
0.2            &     0.06   &       0.06     &  -0.04     &     -0.04                       &  0.2            &    0.20    &      -0.03     &  -0.04     &        1.0                      &   0.2            &     0.20   &     0.33      &     -0.16   &  -0.28                  \\   
0.4            &     0.06   &       0.06     &  -0.006    &     -0.006                      &  0.4            &    0.23    &       0.04     &  -0.004    &        1.0                      &   0.4            &     0.15   &     0.28      &     -0.15   &  -0.27                  \\   
1.0            &     0.05   &       0.05     &  +0.10     &     +0.10                       &  1.0            &    0.29    &       0.10     &  -0.08     &        1.0                      &   1.0            &     0.11   &     0.22      &     -0.9    &  -0.18                  \\   
2.0            &     0.39   &       0.39     &  +0.41     &     +0.41                       &  2.0            &   -0.06    &       0.11     &  +0.16     &        1.0                      &   2.0            &   -0.090   &    -0.09      &     -0.023  &  -0.02                  \\   
\hline

\hline
\end{tabular}
\end{center}
\end{table*}

\section{Results}

Our analysis indicates that the errors on the abundances introduced by
the geometrical distribution of the ionising stars are much smaller
when direct temperature methods are used instead of strong line
methods. We stress here that our aim is not to assess the validity  
of the ICFs but to assess the effect of the distribution of the ionising 
sources on the derived abundances.  Therefore, we are not comparing the 
derived abundances to the 'right answer' (i.e. the input abundances for 
our models), but rather comparing the results from model nebulae with 
distributed ionising sources to those with a centrally concentrated 
source.

Figure~1 shows the logarithmic error, $E_{F-C}$, on the abundances of various
elements due to the spatial distribution of the ionisation
sources. The black solid line shows results using KB94 ICFs and the
red dashed lines shows the results for the ICF scheme of PTR92. 
The errors are due partly to temperature effects (e.g. a steep
temperature profile or differences in the mean temperatures of various
ions) and partly to the ICFs which may be more or less sensitive to
changes in the effective ionisation parameter brought about by a
different spatial distribution of sources (see discussion in
EBS07). Here we are mostly interested in the latter, and in order to
isolate this effect we calculate a correction to
the error due to the ICFs, $\Delta(E_{F-C})$, by comparing the C and F-model theoretical
and empirical ICFs according to the following:   
\begin{equation}
\Delta(E_{F-C}) = log_{10}\left(\frac{ICF_F^O}{ICF_C^O}/\frac{ICF_F^T}{ICF_C^T}\right)
\label{e:corr}
\end{equation}
\noindent where $ICF_F^O$, $ICF_C^O$ are the observational ICFs for
the F- and C-models, respectively, and $ICF_F^T$, $ICF_C^T$ are their
theoretical counterparts calculated by the photoionisation model. 

The logarithmic errors and the ICF corrections are summarised in Table~2.
In the following we will discuss the sources of error in more detail
for each element.

\subsection{Helium} 
KB94 do not include an ICF correction for the unseen neutral helium;
the very small $E_{F-C}$ for KB94 shown in Figure~1 is therefore only
due to the lack of correction and the C- and F-models having different 
amounts of neutral helium. PTR92 on the other hand correct for
neutral helium and it is indeed the ICF correction employed that at
low metallicities is sensitive to changes in the effective ionisation
parameter. The $\Delta(E_{F-C})$ corrections given in Table~2
drastically reduce the $E_{F-C}$ values for PTR92. 

\subsection{Oxygen} The abundance
of oxygen derived by the direct temperature method with a two
temperature description of the medium (DT2T) is not very sensitive to
the geometrical distribution of the ionisation sources (i.e. to the
effective ionisation parameter) for the range of metallicities
discussed here. The largest errors occur for metal-rich regions (twice
solar) and are always below 0.15~dex. We note that both the empirical
and theoretical ICFs for oxygen are roughly unity for all models considered
here. The small error at higher metallicities is due to the steepening
of the temperature profile which is more accentuated for the C-models
(see discussion in EBS07 and Stasi\'nska 1980). This causes the oxygen
abundances to be underestimated in the C-models more than in the
F-models producing the error observed.  

\subsection{Nitrogen}
Nitrogen also shows a similar behaviour with a slightly larger error
(0.21~dex) for metal-rich regions. The nitrogen ICFs are not unity and
some of the error shown in Figure~1 are indeed due to a
different response of the ICFs to the change in the effective
ionisation potential. The ICF correction is $\Delta(E_{F-C})~=~-0.14$
for nitrogen in the Z/Z$_{\odot}$~=~2 case, which brings the nitrogen
$E_{F-C}$ to roughly 0.1~dex. The remainder of the error can again be
ascribed to the steepening of the temperature profile as discussed
above.  

\subsection{Neon}
The situation for neon is more complicated. $\Delta(E_{F-C})$ at
Z/Z$_{\odot}$~=~2 is $\sim$+0.3~dex, which actually increases the
magnitude of the error. The large discrepancy
($E_{F-C})~\sim~0.7$!) is due to the displacement of the Ne$^{2+}$ and
O$^{2+}$ regions which results in a significant difference between
T$_e$(Ne$^{2+}$) and T$_e$(O$^{2+}$) at high Z's. Using T$_e$(O$^{2+}$)
in the determination of Ne$^{2+}$ abundance then
results in the large error. Indeed we find that at Z/Z$_{\odot}$~=~2, 
T$_e$(Ne$^{2+}$)/T$_e$(O$^{2+}$)~=~0.85 for the C-model and 0.93 for
the F-model, the difference has a significant impact on abundances
derived via CELs due to their exponential dependance on
temperature. We calculate the correction due to the difference in the
two temperature regions to be -0.61~dex, which brings the total
$E_{F-C}$ to roughly 0.1~dex, which is comparable to what we found for
oxygen and nitrogen. 

\subsection{Argon}
PTR92 do not employ a correction for Ar$^+$, and the small errors
shown are due to the change in the ionisation structure of F- and
C-models. KB94 do include an ICF for Ar, however this is quite
sensitive to changes to the effective ionisation parameter. The
maximum errors for KB94 are of 0.3~dex at solar metallicities against
a maximum error of 0.1~dex obtained by PTR92 at Z/Z$_{\odot}$~=~0.05. 

\subsection{Sulfur}
Sulfur presents larger problems at low metallicities for both methods
with maximum errors of 0.4~dex at Z/Z$_{\odot}$~=~0.05 for PTR92 and
0.27~dex for KB94 at the same metallicity. Most of the error here can
be ascribed to the different responses of the ICFs to the change of
the effective ionisation parameter. Indeed applying the
$\Delta(E_{F-C})$ corrections given in Table~2 brings $E_{F-C}$ to
values smaller than 0.1~dex both for KB94 and PTR92.

The results obtained for the H-models lie in between those of the F
and C models and are therefore not included in this discussion.

\section{Discussion \& Conclusions}

\begin{figure}
\begin{center}
\includegraphics[width=9cm]{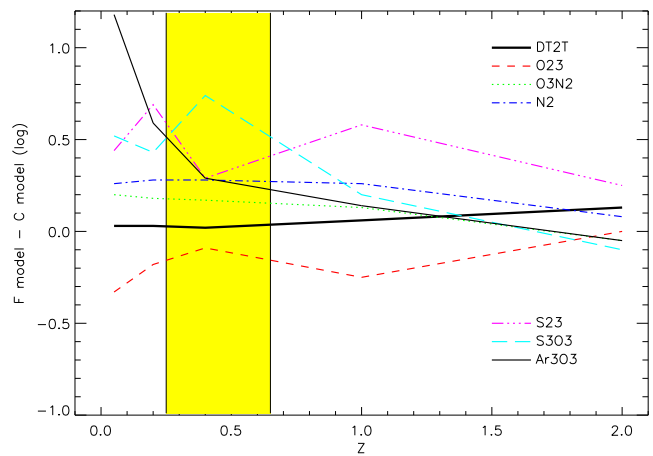}
\caption[]{Errors on the logarithmic abundances calculated via the DT2T
  method caused by the spatial distribution of ionisation sources. The
  black solid line shows results obtained via the DT2T method. The
  thinner lines show results obtained from strong line methods, namely
  as follows. Red dashed: O23 (Pilyugin, 2000, 2001b); 
green dotted: O3N2 (Stasi\'nska 2006); blue dash dotted: N2 (Pettini \&
Pagel 2004); magenta dash double dot: S23 (P\'erez-Montero \& Diaz
2005); cyan long dash: S3O3 (Stasi\'nska 2006); black thin solid:
Ar3O3 (Stasi\'nska 2006). The yellow shaded region indicates the
metallicity range of the H~{\sc ii} regions in NGC~300 analysed by B09.} 
\label{f:f1}
\end{center}
\end{figure}

\begin{figure}
\begin{center}
\includegraphics[width=9cm]{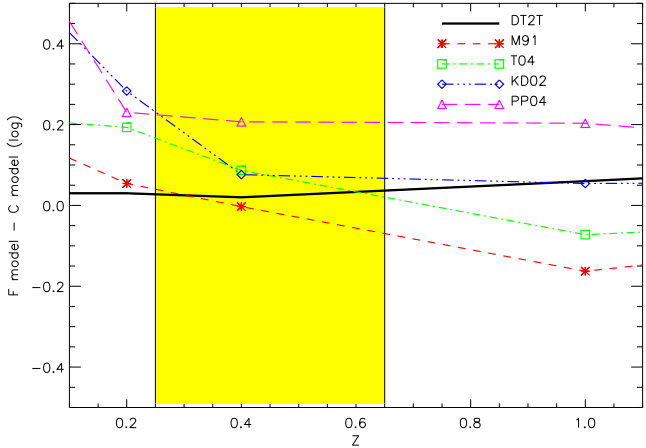}
\caption[]{Errors on the logarithmic abundances calculated via the DT2T
  method caused by the spatial distribution of ionisation sources. The
  black solid line shows results obtained via the DT2T method. The
  thinner lines show results obtained from the strong line methods
  analysed by B09. See text for details. }
\label{f:f1}
\end{center}
\end{figure}

The main conclusion of this short paper is that abundance determinations from
collisionally excited emission lines (CELs) of H~{\sc ii} regions via direct
temperature methods that use a two-temperature description of the
ionised region are very robust and not affected significantly by the
spatial distribution of ionisation sources. 
Indeed the maximum errors on the oxygen abundance derived 
with the DT2T method are still lower than 0.15 dex at Z~=~2 and below
0.05 dex at solar metallicity and below. 

As a comparison we found that the strong-line-methods analysed in
EBS07 gave much larger errors as shown in Figure~2. The figure shows
the discrepancy in the empirically determined oxygen abundance for the
F- (fully distributed) and C- (centrally concentrated) models. The
thick black line indicates the results for the DT2T, while the thinner
lines are for the results of the various strong line methods
considered by EBS07. A detailed legend is provided in the caption to
Figure~2. The small errors associated with the DT2T method compared to
the large errors of the strong line methods are in agreement with the
recent results of Bresolin et al. (2009, B09) who found a systematic
difference between the oxygen abundance calculated by strong line
methods and direct temperature methods for a sample of H~{\sc ii} regions in
NGC~300. The yellow section in Figure~2 shows the metallicity range of
the H~{\sc ii} regions  analysed by B09. B09 compared the metallicities
obtained by a direct temperature and a strong-line analysis of the
emission line spectra of H~{\sc ii} regions in NGC~300 to those obtained from
B and A supergiants in the same galaxy. They found excellent agreement
between the results from supergiants and direct temperature analysis
of the H~{\sc ii} regions, while noticing that a systematic bias affected the
results from some popular calibration of strong lines. The
calibrations used in B09 were not considered by EBS07 and included (i)
the R23\footnote{R23 = ([O II]$\lambda$3727 + [O
    III]$\lambda$37274959,5007)/H$\beta$,} ratio 
from the theoretical calibration of Mc~Gaugh~(1991, M91) using the
analytical prescriptions of Kuzio de Naray et al. (2004) and
Tremonti et al. (2004, their Eq. 1, T04), (ii) the theoretical
prediction for the [N II]$\lambda$6583/[O II]$\lambda$3727 ratio by
Kewley \& Dopita (2002, KD02), and (iii) N2 = log([N
  II]$\lambda$6583/H$\alpha$), calibrated empirically by Pettini \&
Pagel (2004, PP04). 

In order to estimate whether the spatial
distribution of stars may be playing a role in producing the bias
observed by B09 and predicted by ECD07, we have used the emission line
spectra in Table~1 to compute the oxygen abundances given by the M91,
T04, KD02 and PP04 calibrations listed above and compared it to the
DT2T results in Figure~3. The metallicity range of the B09 sample is
again highlighted by the yellow section. The errors of the strong line
methods are comparable to those lamented in the B09 paper, however a
detailed comparison with the observation is premature at this
point. The main problem is that the set of models run by EBS07
comprised a very idealised ionising source population which was
designed to highlight eventual temperature fluctuations that may be
introduced by the distribution of stars with spectra of very different
hardness (see also Section~\ref{s:m}), which turns out to be
equivalent to a much harder 'effective' spectrum than that inferred by
B09 for the H~{\sc ii} regions in NGC~300. The parameter $\eta$ =
(O$^+$/O$^{++}$)/(S$^+$/S$^{++}$), (Vilchez \& Pagel, 1988) was
introduced as a measure of the hardness of the ionising field, with
larger numbers corresponding to a softer spectrum. B09 find an average
log($\eta$) parameter of roughly 0.7, while we find values ranging
between -0.2 and 0.2, indicating a significantly harder spectrum than
that of the B09 H~{\sc ii} regions. Another problem is the fact that EBS07
explored a wide metallicity range and as a consequence the narrow
metallicity range of the H~{\sc ii} 
regions in NGC~300 is very sparsely sampled, as shown in Figure~3 only one
model data point actually falls in that range. In view of these
shortcomings of the models  we can at present only suggest that the
spatial distribution of ionising sources is the cause of the
metallicity bias that afflicts strong line measurements and postpone
firmer statements to a future work where the parameter range is better
suited to match those particular observations.  

We finally note that we have not included a
discussion of the well known abundance discrepancy between CELs and
recombination lines (RLs). A number of possible causes has been
identified in the literature including temperature fluctuations
(Peimbert, 1967), hydrogen-deficient, metal-rich  inclusions (Liu et
al., 2000; Stasinska et al. 2007) and X-ray irradiated quasi-neutral
clumps (Ercolano, 2009). The jury is still out however as to which of
the above effects or a combination thereof is to blame for the
discrepancy. Until the latter problem is resolved all abundances
determined via nebular emission lines carry a potential error. The
excellent agreement between the results obtained by B09 from direct
temperature analysis of CELs and those from the supergiants in
NGC~300, however, indicates that in this Galaxy temperature
fluctuations and X-ray irradiated quasineutral clumps, if at all
present, must be playing a minor role.  

\section*{Acknowledgments}
BE and NB are supported by a Science and Technology
Facility Council Advanced Fellowship. We thank the referee for constructive comments.

\end{document}